\documentclass[12pt]{book}

\usepackage[dvips]{graphicx,color}
\usepackage{makeidx,high}

\makeauthorindex
\makeindex

\begin{document}

\BookTitle{\itshape Extremely High Energy Cosmic Rays}
\CopyRight{\copyright 2002 by Universal Academy Press, Inc.}
\pagenumbering{arabic}

\chapter{
 On the Luminosity  of the Ultra High Energy Cosmic Rays Sources
}

\author{%
Todor STANEV\\
{\it Bartol Research Institute, University of Delaware, Newark, DE19716,
 U.S.A.}\\
stanev@bartol.udel.edu
}


%
\AuthorContents{T.~Stanev} 

\AuthorIndex{Stanev}{T.}

\section*{Abstract}

 The energy density of the Ultra High Energy Cosmic Rays (UHECR) in
 the Universe is a very important parameter for the solution of
 the puzzle of their origin. It defines the luminosity of the UHECR
 sources and thus the type of objects they are.
 This is also of crucial importance for the design of high energy
 neutrino telescopes. The current attempts
 to derive the source luminosity are hindered by the small world
 experimental statistics.
 We show that the unknown strength and structure of the large scale
 cosmic magnetic fields affect strongly the UHECR propagation history.
 The identification of the UHECR sources will bring important information
 on the large scale magnetic fields.  
  
\section{Introduction}
  
  Ever since the discovery of the microwave background and the 
 conclusions about the end of the UHECR spectrum derived by 
 Greisen (1966) and by Zatsepin and Kuzmin (1966), 
 the first 10$^{20}$ eV air shower detected by John Linsley (1963)
 was difficult to interpret. We have not progressed that far in 30 years
 and still argue if the world statistics includes 10 or  20 events.
 Every giant air shower array has registered at least one 
 super-GZK event and now we hope to have more than one order of 
 magnitude increase by the end of the decade. The rational thing to do
 is maybe wait until then to make any conclusions. It is not only the
 intellectual curiosity that makes it very hard to keep silent for such a
 long time. The extragalactic cosmic rays energy density is a
 crucial parameter for the expectations from the fast developing
 high energy neutrino astronomy and for the design of its detectors.
 On top of this we should be better prepared for the analysis
 and interpretation of the forthcoming data.

  The big disappointment of 2002 was the discrepancy between the
 results of HiRes (Abu-Zayyad et al.) in monocular mode and those
 of AGASA (Takeda et al; AGASA web page). One can argue correctly
 that the statistical  significance of the discrepancy is small,
 although such an assessment requires a conspiracy between the
 two groups to  bend their maximal systematic errors in opposite
 directions. 

  There are two types of differences in the measured UHECR spectrum:
 \begin{itemize}
 \item The normalization of the spectrum between 10$^{18.5}$ and 10$^{19.5}$ 
  eV is of the order of the maximum systematic errors of the two
  detection techniques and analyses.\\
 \item The end of the UHECR spectrum is also different. More exactly, the
  HiRes data seem to confirm the GZK feature (Bahcall \& Waxman) while 
  AGASA's do not.
\end{itemize}

\section{The Recent Experimental Data Sets}

\begin{figure}[thb] 
  \begin{center}
    \includegraphics[height=65truemm]{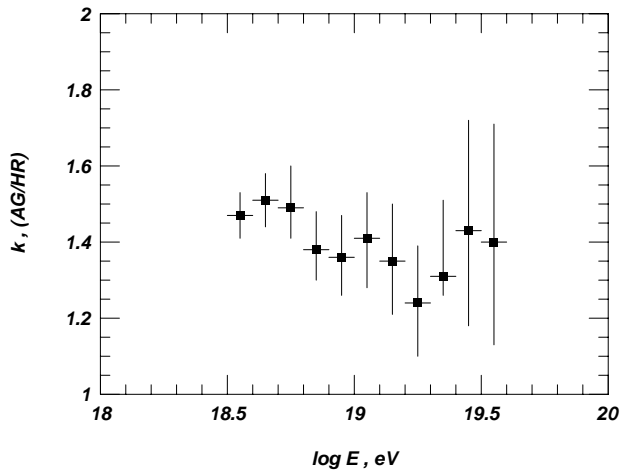}
  \end{center}
\caption{ \protect$k$ factors for the AGASA data based on the HiRes
 normalization. The error bars are calculated using the high/low
 flux estimates of the two experiments.
\label{fig1}
}
\end{figure}

  The two experimental groups have obviously very different energy
 assignments. Since the popular form of the presentation of the spectrum
 is E$^3$ dN/dE the differences are exaggerated in a visual inspection.
 One can however use the data to define better the difference.
 If one experiment assigns the wrong energy $kE$ instead of the
 correct energy $E$,
 \begin{equation}
 (kE)^3 \frac{dN}{d(kE)} = k^2 E^3 \frac{dN}{dE} \; .
 \end{equation}
 The same expression can be used to estimate the difference of 
 the energy estimates of the two experiments without the
 assumption that one of them is wrong. In Fig.~\ref{fig1}
 we show the $k$ parameter derived from the comparison of the
 AGASA and HiRes spectra, which is an indication of the
 difference in  energy assignments as a function of the shower energy.

 The ratio of the the energy assignments $k$ is consistent with a
 constant values of about 1.4 in the whole energy range. Without
 any additional knowledge of the reasons for this difference
 we can draw the conclusion that there are no indications that
 the HiRes energy assignment is influenced by its energy dependent
 aperture.

 Even if we scale the fluxes respectively up(HiRes) and down(AGASA) by
 about 20\% each and eliminate the difference in the normalization,
 the inconsistency in the shape of the spectrum remains, although
 (DeMarco, Blasi \& Olinto) it is of a statistical
 significance lower than 3$\sigma$. Let me speculate for one of the
 possible reasons for the disagreement for the rate of the highest
 energy events. The argument of Bahcall \& Waxman is that HiRes
 has much higher exposure than AGASA but sees one order of magnitude
 less super-GZK events. Indeed, the AGASA experiment gives 
 exposure of 1,460 km$^2$sr.yrs. The exposure of HiRes is more
 difficult to estimate, but from the observational time of 0.275 yrs
 (2410 hrs) of HiRes I and an aperture of 8,000 km$^2$sr one 
 can estimate the exposure as 2,200 km$^2$sr.yrs.

\section{Speculation: Different Fields of View}
 
 There is, however, a big difference in the sky areas that are observed
 by the two experiments. AGASA is restricted to zenith angle of
 45$^\circ$, while the maximum efficiency for the HiRes is at
 higher zenith angles and sensitivity extends up to 80$^\circ$.
 Using a published MonteCarlo zenith angle
 distribution for HiRes (which is in a good agreement with data)
 and assuming a flat zenith angle efficiency for AGASA, I
 estimated viewing efficiency of the two experiments 
 for different regions of the sky.
 The estimate for AGASA is certainly not grossly wrong because of its
 long observation time. The HiRes has only run for a short time and
 has not made its RA distribution uniform, as I have assumed. 
 HiRes' field of view that I estimated should be taken with a grain
 of salt.

\begin{figure}[thb] 
  \begin{center}
    \includegraphics[height=55truemm]{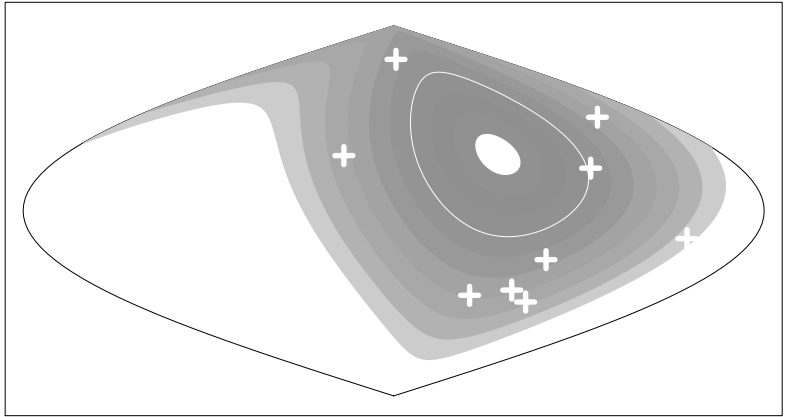}
  \end{center}
\caption{
 The exposure of AGASA and the arrival directions of the showers
 of energy exceeding 100 EeV. The shading is proportional to the
 exposure to different areas in the sky.
\label{fig1a}
}
\end{figure}
 
 Fig.~\ref{fig1a} shows the exposure of AGASA and the arrival
 directions of the super-GZK events. The exposure is calculated
 in declination bands (assuming uniform RA distribution) and then
 plotted in Galactic coordinates. One can outline the region of
 the sky that yields the AGASA events - the white line in
 Fig.~\ref{fig1a}. AGASA has exposure of
 900 km$^2$sr.yrs for this region. HiRes I has a similar exposure
 of 850 km$^2$.sr.yrs. It is certainly premature to claim that
 100 EeV and above events come from certain region of the sky.
 On the other hand, we are searching for the location of their
 sources, and should take into account the different fields
 of view of the experiments.

 \section{Source Luminosity Estimates}

 Returning to the source luminosity estimates,
 the 40\% difference in the energy assignment is one of the
 smallest errors in the determination of the UHECR energy density.
 A much bigger factor is the position at which a researcher
 choses to normalize to the UHECR flux and the
 assumed injection spectrum that is used to fit the data.
 Even for a flat astrophysical bottom-up scenario, a downward shift
 of the normalization point by half an order of magnitude increases
 the luminosity estimate by an order of magnitude in the E$^{-3}$
 part of the spectrum. If steeper injection spectra are considered
 as a better fit to the observed spectrum, the difference could reach
 orders of magnitude. The most important reasons is that
 the total source luminosity should account for the acceleration of
 all lower energy particles that may be hidden behind the Galactic
 cosmic ray spectrum.
 The lowest possible luminosity is predicted for super relativistic shocks,
 where the accelerated particle has a minimum energy $m_p \gamma_{shock}^2$
 (Achterberg et al. 2001), i.e. about 10$^{15}$ eV for $\gamma_{shock}$
 of 1000. In the case of non
 relativistic shocks, where the spectrum extends all the way
 down to the proton mass, the luminosity requirements are higher.

\begin{figure}[thb] 
  \begin{center}
    \includegraphics[height=62truemm]{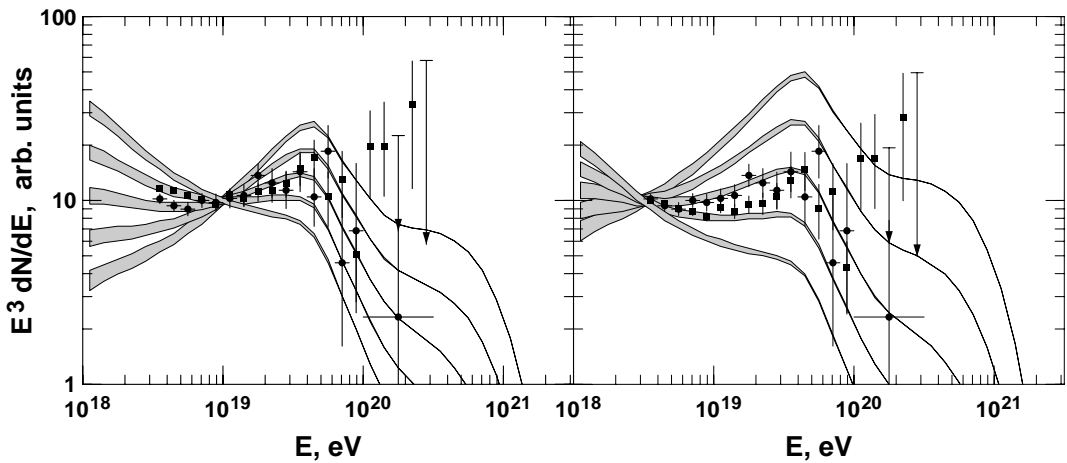}
  \end{center}
\caption{ The data of AGASA (squares) and HiRes (dots) are compared to
 the predictions for the flux arriving at Earth by an isotropic
 source distribution. See text for details. Right hand panel - normalization at
 10\protect$^{18.5}$ eV. Left hand panel - normalization at
 10\protect$^{19}$ eV.
\label{fig2}
}
\end{figure}

 Fig.~\ref{fig2} gives examples of  fitting the UHECR spectrum
 with different injection spectra of isotropic homogeneous source
 distribution neglecting the existence of cosmic magnetic fields.

 The curves in Fig.~\ref{fig2} are for power law injection spectra
 with indeces of 2.00, 2.25, 2.50, 2.75 and 3.00 and an exponential
 cutoff at 10$^{21.5}$ eV.
 The expectations are calculated for two different cosmological
 evolutions of the sources of the form $(1 + z)^n$ with $n$ = 3,4
 (lower and upper edge of shaded spectra) to 
 a maximum at $z_{max}$ = 1.8. The value of $z_{max}$ is irrelevant
 because redshifts larger than 0.5 do not contribute to the fluxes
 above 10$^{18}$ eV for a maximum injection energy of 10$^{21.5}$ eV.

 At least two recent analyses have discussed the data sets in terms
 of the end of the cosmic ray spectrum. Bahcall \& Waxman dismiss
 the AGASA data and reach the conclusion that the GZK cutoff exists
 and is best described by a differential power law injection spectrum
 with $\alpha$ = 2. For injection energies 10$^{15}$ to 10$^{21}$ eV this
 models requires cosmic ray luminosity of 1.4$\times$10$^{45}$
 erg.Mpc$^{-3}$yr$^{-1}$. 

 The visual inspection of the left hand panel of Fig.~\ref{fig2} 
 does not suggest that any of the data sets in question can be fitted
 with $E^{-2}$ injection spectrum. HiRes data seem more consistent 
 with injection spectral index of about 2.5. The same is true for
 the AGASA data up to 10$^{20}$ eV.
 It is worth remembering
 that detailed studies of relativistic shock acceleration predicts
 spectral indeces of 2.2 - 2.3 (Achterberg et al. 2001).
 
 The other analysis (Berezinsky, Gazizov \& Grigorieva), which
 neglects the HiRes results,
 derives an injection spectrum with $\alpha$ = 2.7, that is
 accompanied by top-down origin of the AGASA super-GZK events.
 We have to agree with this conclusion at least in the range 10$^{18.5}$
 - 10$^{19.5}$ eV, where the $\alpha$ = 2.75 injection  predicts
 best the shape of the experimental data. The total luminosity
 required under the same conditions is 4.5$\times$10$^{47}$
 erg.Mpc$^{-3}$yr$^{-1}$. I will use these two numbers to bracket the
 uncertainty in the UHECR luminosity, which is then a factor of 300.

 \section{Cosmic Magnetic Fields}

\begin{figure}[thb] 
  \begin{center}
    \includegraphics[height=65truemm]{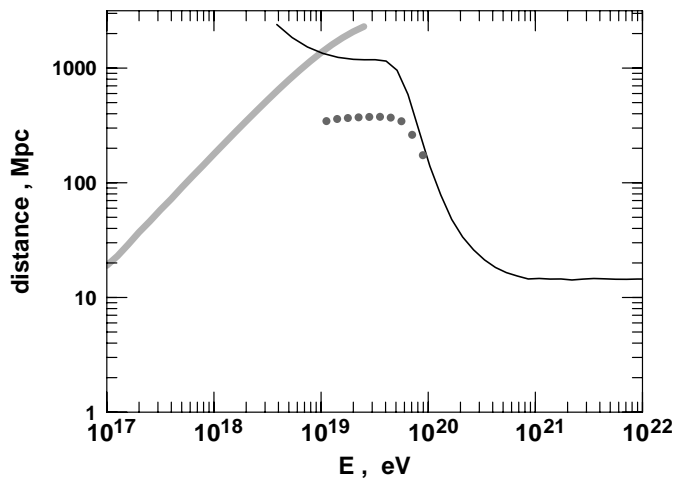}
  \end{center}
\caption{
 Restrictions on the distance from which UHECR can reach us. The solid
 line show the proton energy loss distance. The thick gray line show
 the time delay restriction. The points are for the horizon
 calculated by Stanev et al. (2000).
\label{fig3}
}
\end{figure}

 Fig.~\ref{fig2} demonstrates one potential problem with steep
 injection spectra - such models overproduce at energies around
 10$^{18}$ eV. This excess can be easily accommodated if we 
 account for the cosmic magnetic fields. Achterberg et al (1999)
 derive the scattering angles of UHECR protons in random
 magnetic fields and related increase of pathlength and time
 delay. Assuming small angle scattering, the expression for
 the time delay $\Delta t$ is
\begin{equation}
\Delta t \; = \; 30 \left( \frac{(D/{\rm Mpc})^2
 (B/{\rm nG})^2}{E_{20}^2} \right) (l_0/{\rm Mpc}) \;
 {\rm yrs} \; \; ,
\end{equation}  
 where $E_{20}$ is the proton energy in units of 10$^{20}$ eV and $l_0$
 is the coherence length of the random field.
 This expression does not account for the proton energy loss and is
 the {\em minimum} $\Delta t$.  
 The time delay $\Delta t$ is the excess travel time over the straight
 line propagation time $t$. The total propagation time $t + \Delta t$
 has to be less than Hubble time. This requirement restricts the distance
 from the source to the observer. As an illustration I show on
 Fig.~\ref{fig3} what are the limits on that distance for Hubble time
 of 10$^{10}$ yrs and field strength of 1 nG. 

  Stanev et al. (2000) estimated  the proton energy loss in
 the presence of random magnetic field. The technique applied was 
 Monte Carlo and simulations to Hubble time are very inefficient.
 Fig.~\ref{fig3} shows the maximum distance allowed
 by the Hubble time constraint with no energy loss (thick gray line)
 in 1 nG field and
 the energy loss alone (solid line). The points show a part of the
 transitional region, as calculated for the horizon $R_{50}$ by
 Stanev et al. (2000). Because of energy loss the time constraint
 would be much stronger if the Universe indeed contain
 1  nG random magnetic field and if the cosmic ray sources are 
 isotropically and uniformly distributed. The time delay restriction 
 would eliminate any excess cosmic ray events in the case
 of a relatively steep injection spectrum.
 
  The possible existence of regular large scale fields complicates
 the derivation of the injection spectrum even more. The following 
 exercise by Stanev, Seckel \& Engel (2001) demonstrates the problems:
 a cosmic ray
 source at the origin injects isotropically protons above 10$^{18.5}$ eV
 on a power law spectrum with exponential cutoff at 10$^{21.5}$ eV.
 The source is in the central $yz$ plane of a 3 Mpc wide magnetic wall,
 that is a simplified version of the Supergalactic plane (SGP). 
 Magnetic field with strength of $B_{reg}$ = 10 nG points in $z$ direction
 and decays exponentially outside the SGP. The regular field is accompanied
 by random field with strength $B_{rndm} = B_{reg}/2$.

\begin{figure}[thb] 
  \begin{center}
    \includegraphics[height=65truemm]{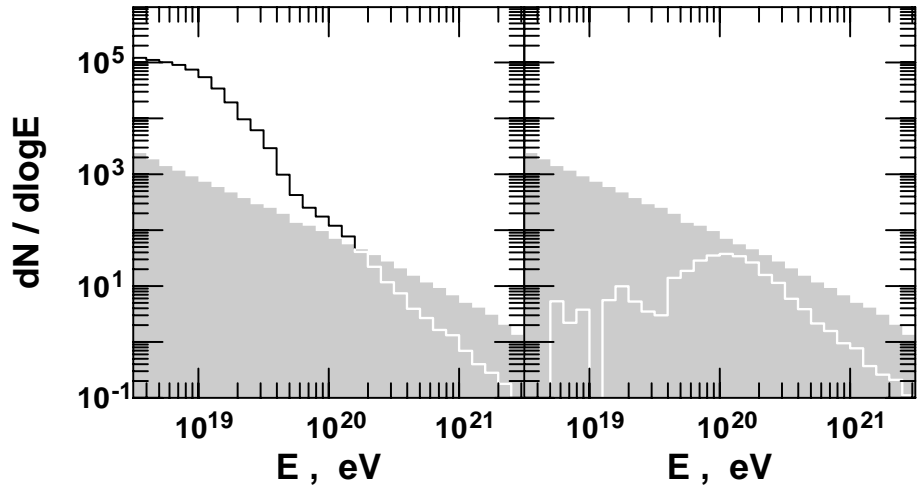}
  \end{center}
\caption{ Energy distribution of the protons leaving the {\em front}
 patch (left) and the {\em side} patch (right) at 20 Mpc from an
 isotropic cosmic ray source. See the text for a description of 
 the geometry. The shaded histogram shows the energy spectra of the
 protons emitted in the direction of the patches.
\label{fig4}
}
\end{figure}

 Protons are followed with energy loss until they intersect a sphere of
 radius 20 Mpc. Their exit positions, velocity vectors and energies are
 recorded. The correlation between these parameters are studied 
 in the analysis of the simulation. Fig.~\ref{fig4} shows
 the energy spectrum of the protons leaving the sphere at two 
 9 Mpc$^2$ patches: the {\em front} patch around $z$ = 20 Mpc inside
 the SGP, and the {\em side} patch with the same area around $x$ = 20 Mpc,
 i.e. in direction perpendicular to the magnetic field.

 The locations of the two patches in Fig.~\ref{fig4} are chosen because
 the exit proton spectra are very different at these positions. 
 Protons of energy below 10$^{20}$ eV are often caught in the 
 SGP magnetic field and can not leave it. They gyrate back and forth
 around the magnetic field lines and are equally likely to leave the
 20 Mpc sphere through the {\em front} and the symmetric {\em back}
 patches. Because of these particles that are trapped in the magnetic
 wall the exit spectra at 10$^{19}$ eV in these patches are higher
 than the  injection spectra by 2 orders of magnitude. At higher energies
 the protons propagate almost rectilinearly. The decrease in the spectrum 
 is due to energy loss.

 Protons exiting through the {\em side} patch show exactly the opposite
 picture. To reach the patch the protons have to cross the magnetic
 field lines and very few lower energy particles can do that with the help of
 the random field. In the vicinity of 10$^{19}$ eV the exit spectrum is 
 more than two orders of magnitude short of the injection spectrum.
 Above 2$\times$10$^{20}$ eV the two exit spectra are identical.

 If two observers were estimating the proton injection spectra 
 with no account for the magnetic fields at 10$^{19}$ eV, their
 estimates would differ by four orders of magnitude. Similar,
 although not as strong, effects are also visible in these patches
 for UHE cosmic ray sources outside the 20 Mpc sphere that illuminate
 the SGP.
 
 In these simple cases one can scale the effects in proton energy
 as a function of the magnetic field strength. If $B_{reg}$ were 5 nG,
 all effects would be the same but at energies that are twice as high.
 Large scale fields of strength 10 nG extending through a small 
 fraction of the volume of the Universe are not an extreme
 assumption. The effects demonstrated in Fig.~\ref{fig4} will certainly
 happen at certain level in the real Universe.

\section{Conclusions}
\begin{itemize}
\item  The energy assignments of the AGASA and HiRes experiments are
 different by about 40\%. This differences appears to be constant between
 10$^{18.5}$ and 10$^{19.5}$ eV. The data of the Auger Observatory in
 hybrid mode should help resolve this difference. 
 The different fields of view of the two
 experiments might also have some relevance to the detected number
 of super--GZK events.
\item  Correct estimates of the UHECR source luminosity are at present
 not possible because of the very limited statistics. All experiments have
 seen super-GZK events but the shape of the spectrum is not well
 determined.
\item  Even in the future, when we hope to increase the available statistics
 by orders of magnitude, this will not be an easy task. The main problem
 is not how high in energy the UHECR spectrum continues, but how low is the
 energy that we have to include in the total source luminosity. The 
 solution should come from the acceleration models. 
 \item This becomes a serious uncertainty if the cosmic ray acceleration
 spectrum is fit with power law spectra steeper than E$^{-2}$. The 
 injection spectrum that fits best the current statistics is not
 flatter than E$^{-2.5}$.
\item  The possible existence of random extragalactic fields restrict
 the distance that protons of fixed energy can reach in Hubble time
 to our local cosmological neighborhood. Extragalactic protons
 below 10$^{17}$ eV are restricted to a few Mpc and those above
 10$^{20.5}$ to about 15 Mpc.
\item  The possible existence of regular fields of extension of 
 $\sim$40 Mpc and strength of order 10 nG affects strongly the propagation
 of 10$^{18}$ - 10$^{20}$ eV protons. The `arrival' spectra in this
 energy range depend on the relative positions of the source
 and the observer with respect to the magnetic field direction and
 structure.
\item Only protons of energy well above 10$^{20}$ eV reveal the source
 spectrum after an account for the energy loss on propagation. Hopefully
 the Auger Observatory, and later EUSO and OWL, will collect significant
 statistics of such events that will reveal the type and the luminosity of 
 the UHECR sources. 
\end{itemize}

\noindent{\bf Acknowledgments} The author is indebted to P.~Sokolsky and
 M.~Teshima for their help in understanding the experimental results
 and for exciting discussions. Much of the work on which this talk is based
 was performed with R.~Engel, T.K.~Gaisser, D.~Seckel and others.
 This research is supported in part by NASA grant NAG5-10919. 

\newpage
\section{References}  
\vspace{\baselineskip} 
\re
1.\  Abu-Zayyad,~T. et al., {\em astro-ph/0208301}
\re
2.\ Achterberg,~A. et al., {\em astro-ph/9907060}
\re
3.\ Achterberg,~A., et al., MNRAS, {\bf 328}, 393 (2001)
\re
4.\ Bahcall,~J.N. \& E.~Waxman, {\em astro-ph/0206217}
\re
5.\ Berezinsky,~V.S., A.Z.~Gazizon \& S.I.~Grigorieva 
 {\em astro-ph/0204357}; {\em astro-ph/0210095} 
\re
6.\  DeMarco,~D., P.~Blasi \& A.V.~Olinto, {\em astro-ph/0301497}
\re
7.\ Greisen, K, Phys. Rev. Lett., {\bf 16}, 748 (1966)
\re
8.\  Linsley,~J., Phys. Rev. Lett., {\bf 10}, 146 (1963)
\re
9.\  Stanev,~T. et al., Phys. Rev. D {\bf 62}:093005 (2000)
\re
10.\  Stanev,~T, D.~Seckel \& R.~Engel, {\em astro-ph/0108338}
\re
11.\  Takeda,~M. et al., Phys. Rev. Lett., {\bf 81}, 1163 (1998) 
\re
12.\  Zatsepin,~G.T. \& V.A.~Kuzmin, Pisma Zh.Exp. Theor. Phys., {\bf 4}, 114
 (1966)
\end{document}